\documentclass[12pt]{iopart}
\usepackage{iopams}
\expandafter\let\csname equation*\endcsname\relax
\expandafter\let\csname endequation*\endcsname\relax
\usepackage{amsmath}
\usepackage{color}
\usepackage{graphicx}
\usepackage{dcolumn}
\usepackage{subfigure}
\usepackage{makecell,tabularx}
\usepackage[normalem]{ulem}
\setcellgapes{3pt}
\makeatletter
\long\def\@makefntext#1{\parindent 1em\noindent 
 \makebox[1em][l]{\footnotesize\rm$\m@th{^{(\arabic{footnote})}}$}%
 \footnotesize\rm #1}
\def\@makefnmark{\hbox{${^{(\arabic{footnote})}}\m@th$}}
\def\@thefnmark{\arabic{footnote}}
\makeatother

\begin{document}

\title{New design of gravitational wave detectors}

\author{Xavier Ja\'en and P. Talavera}

\address{
Department of Physics,
Polytechnic University of Catalonia,\\ Diagonal 647,
Barcelona, 08028, Spain}

\ead{xavier.jaen@upc.edu \& pere.talavera@icc.ub.edu }
\vspace{10pt}

\begin{abstract}
We propose a novel approach to detect gravitational waves based on a semi-rigid detector. The approach relies upon the time delay that the light takes to travel from a fixed mirror at the end of a rigid bar to a nearby free mirror. We show that the dimensions of the experimental device can be shortened in comparison to the conventional ones based on two free mirrors.
\end{abstract}

%
\vspace{2pc}

\section{Motivation}

Gravitational dynamics has been quite quantitatively well tested in the weak-field approximation, i.e. perturbatively close to flat space \cite{Will:2018lln}.  Due to the steadily increasing number of Gravitational Wave (GW) observation from coalescing binaries \cite{LIGOScientific:2018mvr} new techniques to tackle the strong field regime have also been developed \cite{Bern:2019crd}. Besides these efforts and achievements there are some  fundamental issues that are left aside in these developments \cite{barsi}. For instance, although the emission of GW  is a relativistic effect,  its detection could be understood as non-relativistic, depending on the way we use light to measure distances.

Let's start with a simple exercise. In the detection zone, we can find  a vacuum  solution $\phi$ for the non-relativistic Poisson equation $\Delta \phi  =  4\pi G\rho$ in the form 
\begin{equation}\label{eqpotnewt}
\phi  = {\phi _ \times }(t) \,x\, y + {\phi _ + }(t)({x^2} - {y^2})\,,
\end{equation}
where $\phi _ \times$ and  $\phi _ +$ are two arbitrary functions that could be related to the graviton polarization.

The equation of motion for test particles can be derived from the usual Newtonian Lagrangian
\begin{equation}\label{lagnewtphi}
L=\frac{1}{2} m\dot{\vec x}^{\,2}-m\phi\,,\quad \vec x=(x,y,z)\,.
\end{equation}
Henceforth dots will stand for derivatives with respect to time.
Working up to linear order in $\phi$,  with the initial data $(x_0,y_0,z_0=0)$, the test particles trajectories at rest when $\phi=0$ take the form
\begin{equation}\label{eqtest}
\begin{array}{l}
x = {x_0} + \frac{1}{2}\left( {{h_ + }(t) {x_0} + {h_ \times }(t) {y_0}} \right)\,, \\
y = {y_0} + \frac{1}{2}\left( {{h_ \times }(t) {x_0} - {h_ + }(t) {y_0}} \right)\,, \\
z = 0\,,
\end{array}
\end{equation}
with
\begin{equation}
h_\times=-2\int^t_0\left(\int^t_0 \phi_\times dt\right)dt\quad \mbox{and}\quad h_+=-4\int^t_0\left(\int^t_0 \phi_+ dt\right)dt\,.
\end{equation}
It is interesting to notice that one can write a rigid covariant form equivalent to the Lagrangian (\ref{lagnewtphi}) as
\begin{equation}\label{lagnewtV}
L^\prime = \frac{1}{2}m (\dot{\vec x} - \vec V)^2\,,
\end{equation}
being $\vec V=(V_x,V_y,V_z)$ any velocity field for the test particles trajectories solutions of the equations of motion derived from (\ref{lagnewtphi}) \cite{Jaen:2015}. From (\ref{eqtest}), and working up to linear order in $h_\times$ and $h_+$, we can choose $\vec V$ as
\begin{equation}\label{Vpot}
V_x =\frac{1}{2}\left({{\dot h_ + } x + {\dot h_ \times } y}\right)\,, V_y = \frac{1}{2}\left({{\dot h_ \times } x - {\dot h_ + } y}\right)\,, V_z = 0\,,
\end{equation}
and write the linearised version of (\ref{lagnewtV}) as
\begin{equation}\label{lagnewtlinearV}
L^\prime =  \frac{1}{2}m \left(\dot{\vec x}^{\,2}-2\vec V\cdot \dot{\vec x}\right)\,,
\end{equation}
in which we want to point out that it does not contain an independent term in $\dot{\vec x}$.

In a suitable reference system  we can interpret (\ref{eqpotnewt}) as the Newtonian potential of a linear plane GW. That is, the  expression (\ref{eqtest}) is just identical to that derived in GR within the linear approximation in the metric perturbation, restricted to zero order in $c^{-1}$ and taking $t$ as the proper time at the origin \cite{Bondi:1958aj}. Eq. (\ref{eqtest}) takes into account how a free test mass is influenced when a plane linear GW interact with it. If this is so, to what extend are we not merely measuring Newtonian effects on test particles? \cite{Press:1972am}. 
If, with a sufficient accuracy, we identify the coordinates $(x, y)$ in (\ref{eqtest}) with the points of a real rigid body and $t$ with the proper time of each of these points, the detection of a GW could be reduced to a Newtonian measurement.

In sections 2 and 3 we shall discuss to what extent this is possible using the rigid gauge (RG) which implies a particular type of Fermi coordinates, the Fermi-Rigid (FR) coordinates. At the required order FR coincides with the local Lorentz (LL) gauge  which, in turn, is usually identified with the lab reference system (LAB).

The currents experiments to detect GW, as LIGO, are based on the detection of the motion of free falling particles. 
Because the Gaussian (GG gauge) are directly identified with the motion of free particles this is the most natural choice for the coordinates. The transverse-traceless (TT)  gauge, widely used in the study of linear plane GW, is a particular case of the GG gauge. 

Beyond the LIGO like configurations, but still using interferometry, it is legitimate to ask whether the introduction of some kind of rigid element in the experimental setup can improve, in some aspect, the detector performance. We are neither referring to detectors based on the resonance of a more or less rigid objects, as in Weber like configurations \cite{weber} nor to the optical rigid bars configuration as in \cite{Braginsky:1997}. We refer to the detection of the relative motion between two independent mirrors, in the line of the LIGO like devices, 
and whether the introduction of some kind of rigid element can cause an improvement of the device attributable exclusively to the rigidity property and not from the presumed lack of rigidity, which effectively could  cause resonance phenomena. In sections 4 and 5 we analyze this possibility. In sections 6 and 7 we study its feasibility from the experimental point of view. In section 8 we briefly point out two possible configurations of the entire detector. Finally, in section 9, we analyze how the various noises could affect the main parts of the detector.

\section{General Relativity in two gauge}

Due to general covariance, the metric in a general space-time   can be parametrized in terms of ten potentials. Out of these only six are independent once coordinate transformations are taken into account. We shall highlight two formulations that directly use these. 

{\bf The Gaussian gauge}: 
the widest known form of GR is the Gaussian gauge.  We can always describe the proper time and space coordinates of a given congruence of free particles as $\left\{ T,X^i=\right.$constant$\left.\right\}$. Within this coordinate system the metric can be cast  as
\begin{equation}\label{metG}
d{{\cal T}^2} = d{T^2} -  \textstyle{\frac{1}{{{c^2}}}}{g_{ij}}d{X^i}d{X^j} \,,  
\end{equation}
where $g_{ij}$ are the six independent potentials\footnote{The latin indices run from  $1$ to $3$.}. Thus an observer can measure the motion of any particle by referring it to the given congruence of \textit{moving clocks}.
As we have already mentioned, the transverse-traceless gauge is a particular case of the GG gauge. Notice that the GG gauge is not suitable to describe rigid motions nor to describe the LAB system. For this, we need a separate input that defines \textit{rigidity}. 

An analogous Newtonian formulation can be obtained considering  $c\to\infty$ for the test particle and field equations derived from (\ref{metG}) \cite{Lifshitz:1980}. This Newtonian formulation could be useful to study cosmological problems.

{\bf The Fermi-rigid gauge}: as Fermi showed \cite{Fermi:1922}, along any time-like geodesic, with proper time $t$, we can choose a set of space-time coordinates such that the metric coincides with Minkowski spacetime up to quadratic order in the space coordinates. There are many exact space-time coordinate systems that led to the above Fermi condition. In order to choose one in particular Manasse and Misner \cite{Manasse:1963} introduced the Fermi Normal (FN) coordinates. These have the advantage of having a nice geometrical interpretation in terms of  geodesic distances\footnote{There is still some controversy on the subject, see for instance  \cite{Rakhmanov:2004eh,Mazlin:1994} and especially \cite{Delva:2012}.}.

In \cite{Jaen:2018wpg} it was shown that  one can simultaneously:  
\begin{enumerate}
\item Describe a chosen time-like geodesic with proper time and position by $(\bar t,0)$. 
\item Surround it with a geodesic congruence with velocity field $V^i(\bar t,{x^j})$, of at least order ${\cal O}(x)$, and proper time field fulfilling  $\tau (\bar t,{x^j})=\bar t+{\cal O}(x^2)$.
\item Write the metric as
\begin{equation}
\label{metRGR}
d{{\cal T}^2} =   d{\tau ^2} -\textstyle{\frac{1}{{{c^2}}}}\left( \tau _{,\,i} \tau _{,\,j} + \gamma _{ij} \right)(d{x^i} - {V^i}d\bar t)(d{x^j} - {V^j}d\bar t )
\end{equation}  
where, as usual, $f_{,i}\equiv \frac{\partial f}{\partial x^i}$. In addition we defined  ${\gamma _{ij}} := \frac{1}{{\cal H}^2}{\delta _{ij}} - {\sigma _{,\,{i}}}{\sigma _{,\,{j}}}$\,, being $\sigma$ of at least order ${\cal O}(x^2)$ and for the cases in which we are interested in  one can set ${\cal H}=1$. This form of the metric will be called the rigid gauge, RG, related to the coordinates $(\bar t,{x^j})$. For the RG observer, test particles near the origin move without acceleration with respect to the clocks at rest.
\item  The coordinates $(t,x^i)$, with $t=\tau (\bar t,{x^j})$, will be a particular case of Fermi coordinates, henceforth dubbed Fermi-rigid (FR) gauge.
\end{enumerate}

The six independent potentials of the RG gauge in (\ref{metRGR}) are $\tau\,, V^i\,, {\cal H}$ and $\sigma$. Notice that (\ref{metRGR}) is covariant under the usual group of rigid motions parametrized with $\bar t$. This formulation aims to make possible:
\begin{enumerate}
\item  To measure the distance $\Delta\ell$ between two events, $(t_1,x_1^i)$ and $(t_2,x_2^i)$, with $t_1=\tau(\bar t,x_1^i)$ and $t_2=\tau(\bar t,x_2^i)$, using a nearby rigid body at rest.
\item The result of this procedure coincides with the Euclidean distance.
\end{enumerate}
When the FR gauge, $(t,{x^j})$ coordinates, is used up to certain order it matches the local Lorentz gauge. 

As in the case of the GG gauge, the standard Newtonian formulation can be obtained from both RG and FR gauge considering  $c\to\infty$ .

\section{Gravitational waves in the detection zone}

For the time been we focus in the $h_+$ mode contribution to the GW. In the TT gauge, with $X^i=\left\{ {X,Y,Z} \right\}$,  the metric 
for a linear plane wave traveling in the $Z$ direction is
\begin{equation}
\label{coffZ}\hspace{-10mm}
d{{\cal T}^2} =  d{T^2} - \textstyle{\frac{1}{{{c^2}}}}\left( {\left\{ {1 + h_+( T - \textstyle{\frac{Z}{c}})} \right\}d{X^2} + \left\{ {1 - h_+( T - \textstyle{\frac{Z}{c}})} \right\}d{Y^2} + d{Z^2}} \right)\,.
\end{equation}
The GW detector will be located at the detection zone, $Z = {\cal O}(h_+)$, and will be at rest when no perturbation is present. Thus when studying either the trajectories of photons or the mirrors one can neglect the motion perpendicular to the detection plane. 
In addition, for the photons, deviations of order $h_+$ in the trajectory will give terms of order $h_+^2$ and can be safely neglected.  This means that we can apply the $h_+$-linearity and take $h_+\left(T-\frac{Z}{c}\right)\to h_+\left( {T} \right)$ in (\ref{coffZ}), thus giving 
\begin{equation}
\label{coff}\hspace{-10mm}
d{{\cal T}^2} =  d{T^2} - \textstyle{\frac{1}{{{c^2}}}}\left( {\left\{ {1 + h_+( T)} \right\}d{X^2} + \left\{ {1 - h_+( T )} \right\}d{Y^2} + d{Z^2}} \right)\,.
\end{equation}
This will be our departing metric. It takes the RG gauge ($ \bar t,x^i$ coordinates) form (\ref{metRGR}) changing spatial coordinates to
\begin{equation}\label{Yx}
X = x - \frac{1}{2}h_+(\bar t) x\,,\qquad Y = y+ \frac{1}{2}h_+(\bar t)y\,,\qquad Z=z\,,\qquad T=\bar t\,.
\end{equation} 
Notice that the spatial part of this change is suggested by the Newtonian expression (\ref{eqtest}) with $t\rightarrow\bar t$, $h_\times\rightarrow 0$ and $(x_0,y_0,z_0)\rightarrow (X,Y,Z)$.
Following \cite{Jaen:2018wpg} we can identify the right potential $\tau(\bar t,x^i)$, present in the metric (\ref{metRGR}), for which (\ref{coff}) takes the form $d{{\cal T}^2} =d\tau^2-\frac{1}{{{c^2}}} d\vec x^{\, 2}+  {\cal O}(x^2)$
\begin{equation}\label{Tt}
\tau(\bar t,x^i)= \bar t + \frac{1}{{4{c^2}}}\dot {h}(\bar t)\left( {{x^2} - {y^2}} \right)\,,
\end{equation} 
where dots stand for the derivatives with respect to the argument.

It is also worth noticing that the Newtonian Lagrangian derived from the RG gauge is exactly (\ref{lagnewtlinearV}) once (\ref{Vpot}) and $h_\times=0$ have been taken into account. Once we have identified the rigid motions, using the RG gauge,  we can interpret the $x^i$ coordinates as the points attached to a suitable rigid body even for relatively long $(x, y,z=0)$ distances from the origin, with the obvious limitation that the body maintains its rigid condition.

The FR gauge ($t,x^i$ coordinates) can be found after identifying the time coordinate $t$ using (\ref{Tt}): $t=\tau(\bar t, x^i)$. The explicit form of the metric after these changes is
\begin{equation}
\label{metr}
d{{\cal T}^2} = \left( {1 + \frac{{2\phi }}{{{c^2}}}} \right)d{t^2} - \frac{1}{{{c^2}}}\left( {d{x^2} + d{y^2} + d{z^2}} \right),
\end{equation}
with $\phi  =  - \frac{1}{4}\ddot {h}_+(t)\left( {{x^2} - {y^2}} \right)$. (\ref{metr}) coincides with the usual LL gauge when neglecting normal motions to the detection plane. 

For the calculations that we will present there are several advantages to consider (\ref{coff}) and (\ref{metr}) together with (\ref{Yx}). Both allow us to omit from the beginning the motions of photons and mirrors normal to the detection plane. These motions will be described as $Z=z=0$. Besides that, (\ref{coff}) and (\ref{metr}) with (\ref{Yx}) describe the same mathematical object linear in $h_+$. Of course the reasonings about the location of the rigid points only applies to the  $x$ and $y$ coordinates and this is just what we need. Note that no order in $c^{-1}$ reasonings have been used. This means that we are free to use either one of the two expressions in the same detection plane and that the results only have the restriction of linearity in $h_+$, but not in the orders in $c^{-1}$ or how far we are from the origin.

Notice that $t$ is the proper time at the origin $x^i=0$, and can be identified with the TT gauge time $T$ at the same point. The space coordinates for particle geodesics at rest in the detection plane when $h_+=0$ are, from (\ref{Yx}),
\begin{equation}\label{xyzG}
x = x_0 + \frac{1}{2}h_+(t) x_0,\qquad y = y_0- \frac{1}{2}h_+(t)y_0,\qquad z=0\,.
\end{equation}

Finally, from (\ref{Yx}), setting $x=X_0$ and  $y=Y_0$  we can describe the rigid  locations $(x,y)$ at the detection plane in TT coordinates, as
\begin{equation}\label{XYZR}
X = X_0 - \frac{1}{2}h_+(T) X_0,\qquad Y = Y_0+ \frac{1}{2}h_+(T)Y_0,\qquad Z=0\,. 
\end{equation}
In what follows and for the sake of clarity, we will call always TT the coordinates or gauge used in (\ref{coff}) and LL the coordinates or gauge used in (\ref{metr}).

\section{Effects of gravitational waves on the time delay: three different setups}

Although observables are gauge independent quantities, we expect a set of coordinates to be more suitable than others to parametrize a given physical observable. For instance, in the case of the TT gauge, the coordinates are given by the positions and times of the free falling clocks. When interacting with a GW those positions oscillate with respect to the LL gauge, but by definition they remain fixed within the TT gauge. It is obvious that the use of rigid bars in the experimental device is better described in the LL gauge.

To simplify the physical setup we shall consider a $(1+1)$ spacetime and use both the TT, $(T,X)$, and the LL, $(t,x)$  gauges, (\ref{coff}) and (\ref{metr}) respectively, denoted generically by $(\lambda,z)$ coordinates. In addition we assume a GW with the plus polarization coming from the $z$-direction. We shall follow the standard analysis \cite{Maggiore:2008} and calculate how a GW affects the propagation of light between the end points of an interferometer. For illustrative purposes we consider a single arm detector. We are after the round trip time delay, $\delta(\lambda)$, of a photon traveling between two mirrors. Notice that either time $T$ or $t$ will be measured by the same clock at the origin. Consider a photon leaving a mirror $A$ located at $z_A$ at time $\lambda$ and arriving at the mirror $B$ located at $z_B$\footnote{We consider in what follows $z_B>z_A$. }  (forward direction, $z_+$). If there would be not GW distortion the traveling time interval between the two mirrors would be $\frac{1}{c}({z_{B0}} - {z_{A0}})$, being $z_{A0}$ and $z_{B0}$ the location of the mirrors when no GW is present, $h_+=0$. Because the interaction with the GW, this time interval must be corrected with terms linear in $h_+$. Thus we expect a traveling time lapse ${\Delta _ + }\lambda-\frac{1}{c}({z_{B0}} - {z_{A0}})$. The photon will bounce and travel in the reverse direction (backward $z_-$) reaching the initial mirror. The lapse time taken in this second trip is given analogously by ${\Delta _ -}\lambda-\frac{1}{c}({z_{B0}} - {z_{A0}})$. The previous picture dictates the boundary conditions to be fulfilled by the photon trajectories, $ z_\pm(\lambda)$, at each mirror
\begin{equation}
\label{PHgencond}
\begin{array}{*{20}{l}}
{{\mbox{forward:}}\left\{ {\begin{array}{*{20}{l}}
\mbox{Initial}\,\,\,{{z_ + }(\lambda ) = {z_A}(\lambda )}\,,\\
\mbox{Final}\,\,\,{{z_ + }(\lambda  + {\Delta _ + }\lambda ) = {z_B}(\lambda  + {\Delta _ + }\lambda )}\,.
\end{array}} \right.}\\
{}\\
{{\mbox{backward:}}\left\{ {\begin{array}{*{20}{l}}
\mbox{Initial}\,\,\,{{z_ - }(\lambda  + {\Delta _ + }\lambda ) = {z_B}(\lambda  + {\Delta _ + }\lambda )}\,,\\
\mbox{Final}\,\,\,{{z_ - }(\lambda  + {\Delta _ + }\lambda  + {\Delta _ - }\lambda ) = {z_A}(\lambda  + {\Delta _ + }\lambda  + {\Delta _ - }\lambda )\,.}
\end{array}} \right.}
\end{array}
\end{equation}
Bearing in mind that photons travel along null geodesics, one can integrate $d{{\cal T}^2}=0$ to obtain, up to order $h_+$, the photon trajectory in each system of coordinates 
\begin{enumerate}
\item[(a)] In TT coordinates
\begin{equation}
\label{photontrag}
{X_{\pm }}(T) = \pm c \left( T   - \frac{1}{2}  \int_0^T dT^\prime h_+(T^\prime) \right)+  K_ \pm \,.
\end{equation}
\item[(b)] In LL coordinates
\begin{eqnarray}
\label{photontrar}
\hspace{-10mm}
{x_{\pm }}(t) =  &&\pm c\,  t +  k_ \pm - \frac{1}{4 c} \left\{ c^2 \left[ 
\dot h_+(t) t^2 - 2 t\, h_+(t) + 2 \int_0^t dt^\prime h_+(t^\prime)\right]  \right.\nonumber \\ 
&& \left.    \mp  2 c\, k_ \pm \left[  h_+(t)-h_+(0) -t\, \dot h_+(t)  \right]
+ k_ \pm^2  \left[ \dot h_+(t)-\dot h_+(0)  \right] \frac{}{}\right\} \,,
\end{eqnarray}
\end{enumerate}
being $K_ \pm$ and $k_ \pm$  integration constants which can include a term of  order $h_+$. 
If we impose to (\ref{photontrag})-(\ref{photontrar}) the boundary conditions (\ref{PHgencond}) we can identify the time delay
\begin{equation}
\label{zdelay}
\delta (\lambda ) = {\Delta _ + }\lambda  + {\Delta _ - }\lambda  - \frac{2}{c}({z_{B0}} - {z_{A0}})\,.
\end{equation}
In the absence of GW $\delta(\lambda)=0$  and the combination $\Delta_+ \lambda+\Delta_-\lambda$ reduces to the flat photon round trip time.

Before getting into details, we want to stress where we stand and where we want to get.
So far we have discussed two types of coordinates: the TT gauge and the LL gauge.  Given an observable we 
want to stress that any of such coordinates will lead to the very same result for it. This is neither more nor less than diffeomorphism invariance.

We have engineered three different observables. Although at first sight they seem almost identical they are indeed different: the time delay we shall refer corresponds to different physical situations, depending on the boundary conditions, of the mirrors. Some of these setup are more easy to tackle in one of system of coordinate than in the other, but as argued above the final result is independent of this choice. Let's add that, since our observables are time intervals measured by a clock at the origin, these intervals can be identified by the intervals of the  times $T=t$ at the same point.

 \begin{figure}
\centering
\subfigure[]{
\includegraphics[scale=0.2]{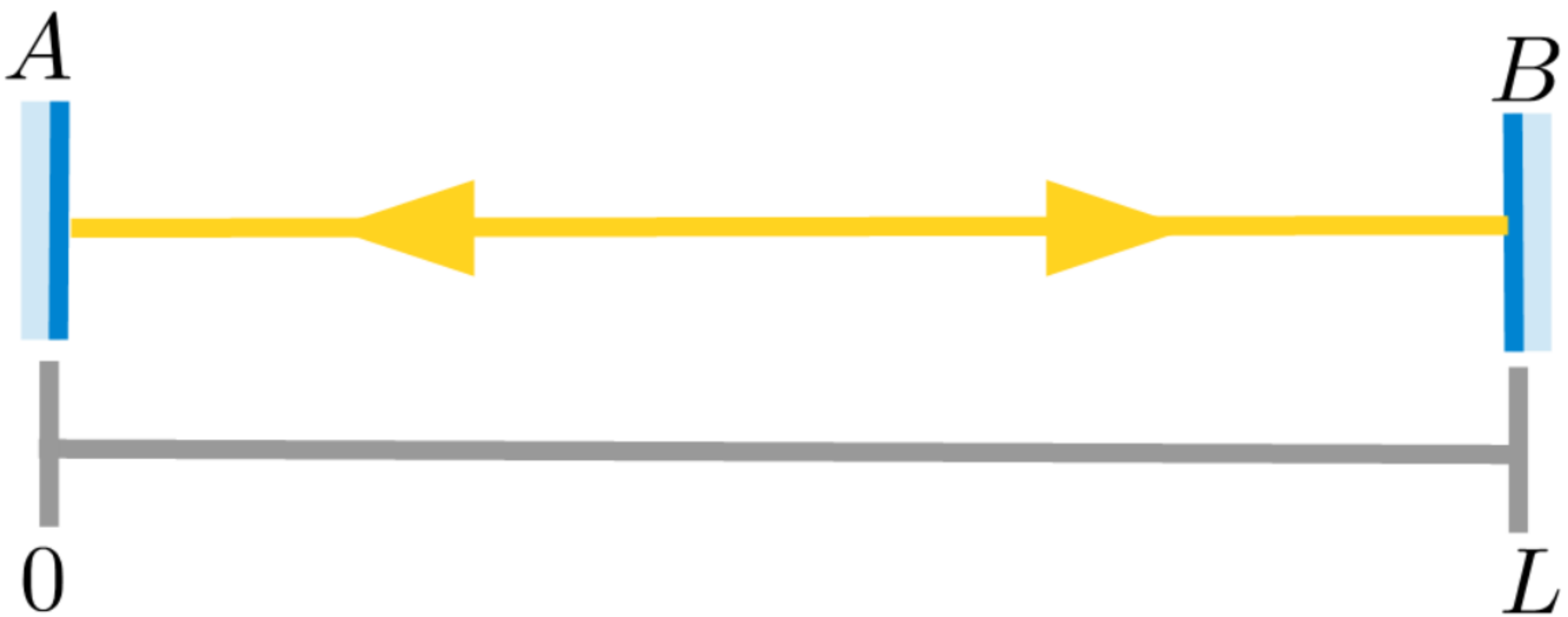}}
\subfigure[]{
\includegraphics[scale=0.2]{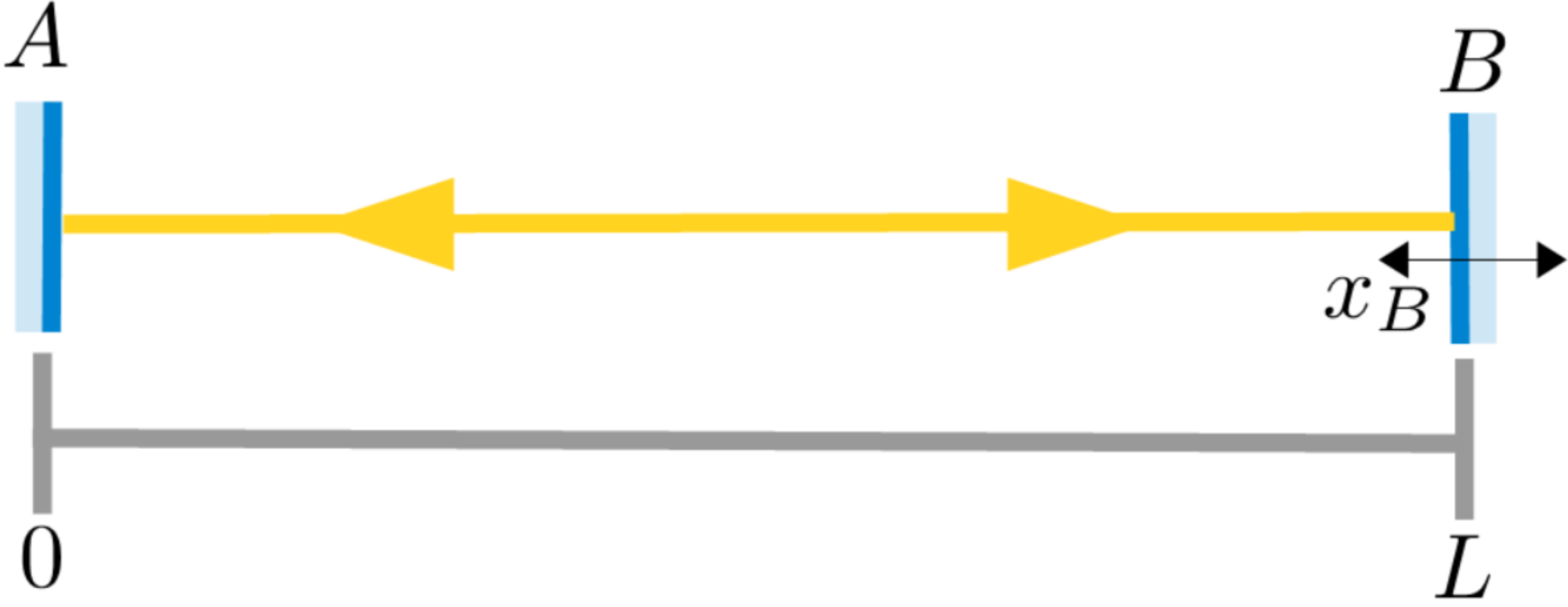}}
\caption{A LIGO like device (LG) consisting of two free mirrors $A$ and $B$ located at $ X_A = 0 $ and $ X_B = L $  in TT coordinates (a) and at $ x_A = 0 $ and $x_B = L + \frac{1}{2}h_+(t)L$ in  LL coordinates (b).}
\label{figLIGODevice}
\end{figure}

\begin{enumerate}
\item[(LG)] {\bf Delay time for two free-falling mirrors: a LIGO like device}

For LIGO the mirror $A$ is located at the origin, which could be understood as both geodesic and rigid, i.e. $X_A=x_A=0$. The mirror $B$ is free falling. If we describe it in TT coordinates its position remains at $X_B=L$ even as the GW pass through it. If instead we choose  LL coordinates the mirror $B$ oscillates with respect to this system, $x_B = L + \frac{1}{2}h_+(t)L$,  see Fig.~\ref{figLIGODevice}.

Either using (\ref{photontrag}) or (\ref{photontrar}) we find that (\ref{zdelay}) gives
\begin{eqnarray}
\label{delayI}
\delta^{\rm LG} (t,L)  = \frac{1}{2}\int\limits_{t}^{t + \frac{{2L}}{c}} {h_+(t^\prime)\,} dt^\prime
\,,
\end{eqnarray}
in agreement with \cite{Rakhmanov:2004eh}. 

\item[ (PR)] {\bf Delay time for an optical bar: rigid device}

Let's stop by in another extreme situation that should clarify our claim: a rigid bar of length $2R$, with its midpoint located at the origin. Both, the midpoint and one of the ends incorporate assembled a mirror. The mirror on the midpoint follows a geodesic trajectory but the other mirror is at fixed rigid distance $L$ from the origin. It is obvious that the most suitable set of coordinates to describe any observable is the LL. 
An experimental setup with this configuration was already implemented in \cite{Braginsky:1997}, see Fig.~\ref{figSRLD}-(a).

Solving (\ref{photontrar}) under conditions (\ref{PHgencond}), with $\lambda\to t$, $z\to x$, $x_A=0$ and $x_B=L$, one obtains for the time delay
\begin{eqnarray}
\label{delayIII}
\delta^{\rm PR}(t,L)  = \left(1-\frac{\omega L}{c}\csc\left(\frac{\omega L}{c}\right)\right) \delta^{\rm LG}(t,L)\,.
\end{eqnarray}
We can interpret this result as the measure of the radar-length variations of a rigid body due to the interaction with a GW.

\begin{figure}
\centering
\subfigure[]{
\includegraphics[scale=0.2]{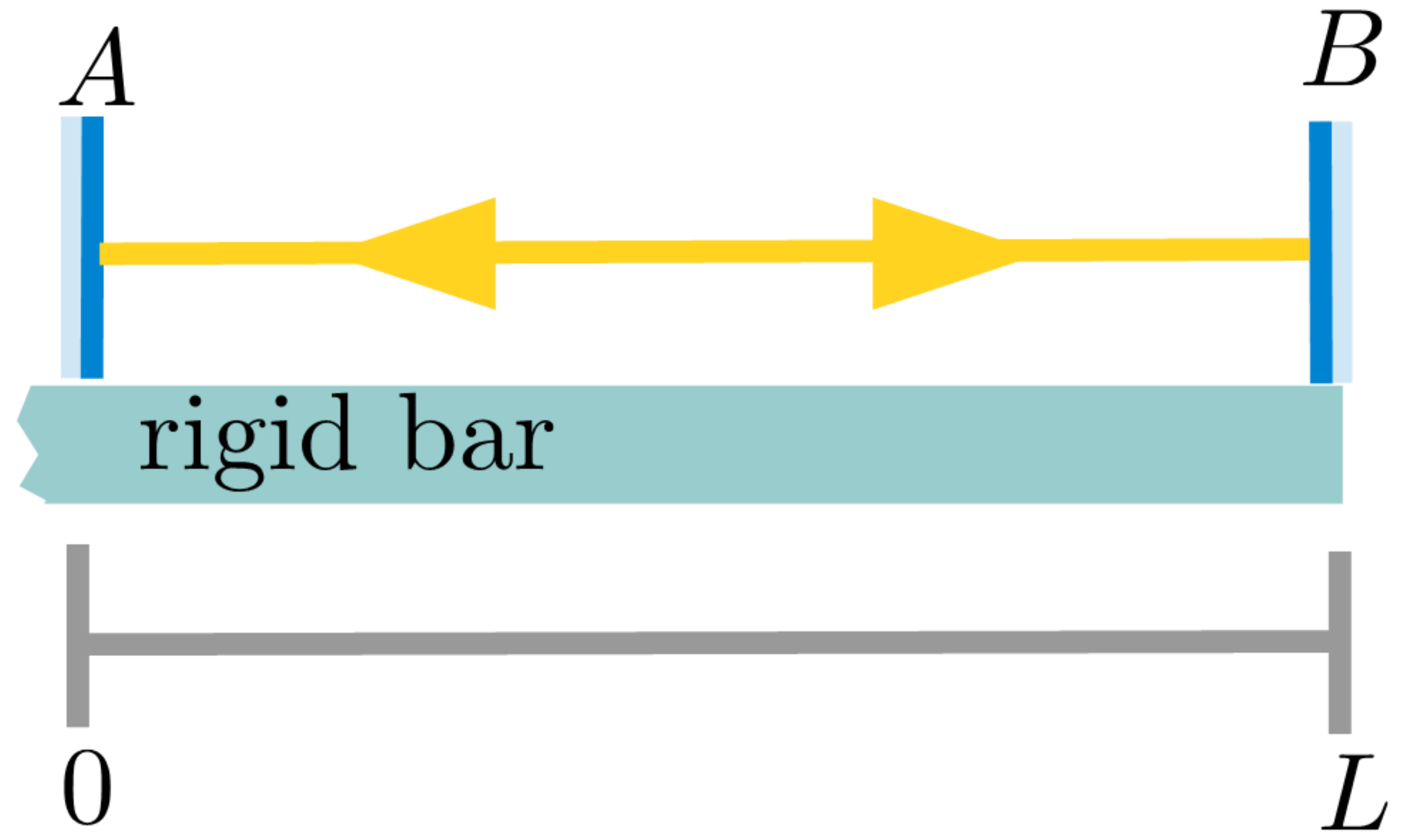}}
\subfigure[]{
\includegraphics[scale=0.2]{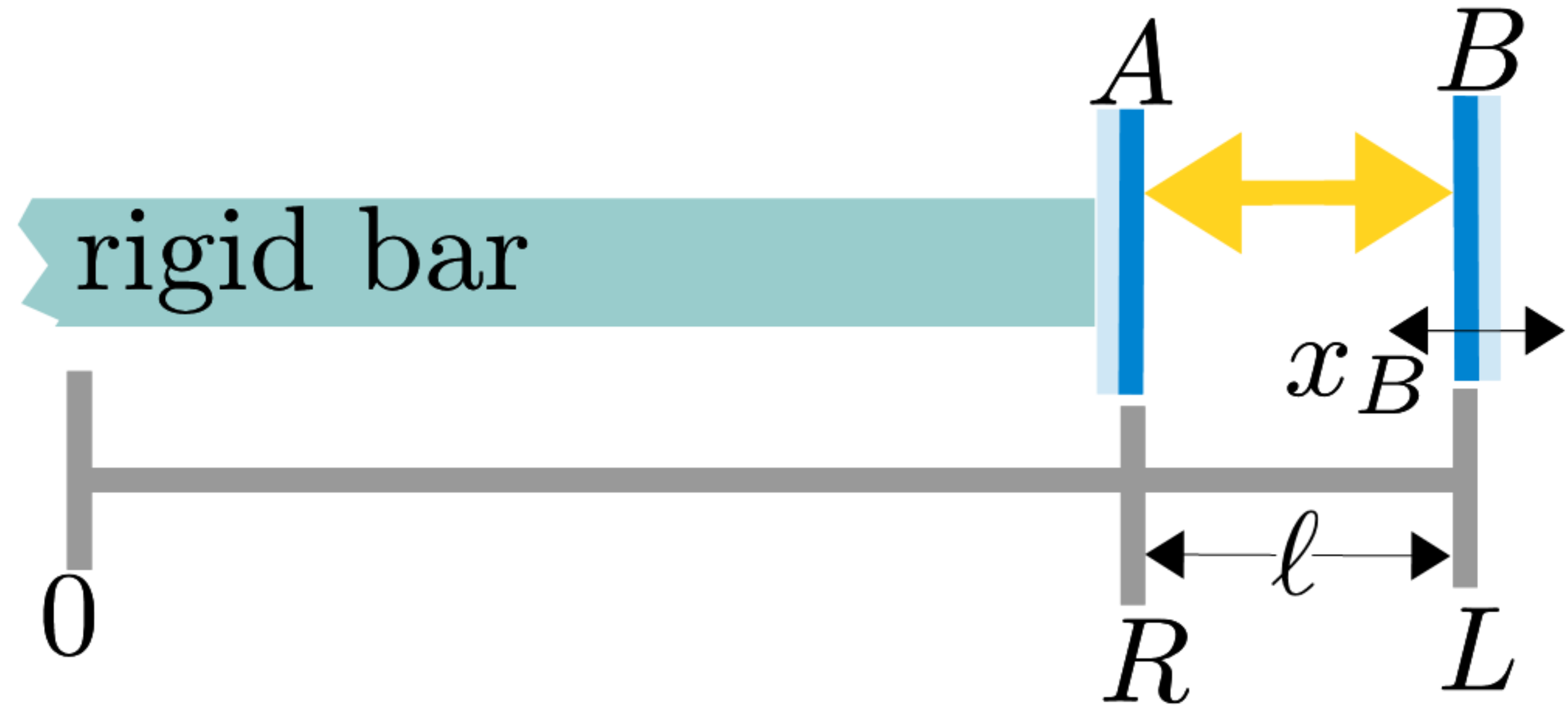}}
\caption{(a) A pure rigid type device (PR) consisting of two mirrors: $A$ free falling at $ x_A = 0 $ and $B$ fixed at $x_B=L$. The rigid bar has length $2L$. (b) A semi-rigid type device (SR) consisting of two mirrors: $A$ fixed at $ x_A = R $ and $B$ free at ${x_B}= L + \frac{1}{2}h_+(t)L$. The rigid bar has length $2R$. We ignore the delay time due to the $[0,R]$ strip. In both cases we use LL coordinates.}
\label{figSRLD}
\end{figure}

\item[ (SR)] {\bf Delay time for a semi-rigid type device}

The next and last observable is to the best of our knowledge a new proposal and under some circumstances, to be described below, is the most interesting case. Is a merging of the previous two situations, see Fig.~\ref{figSRLD}-(b). The setup consist of two elements: {\sl i)} a rigid bar of length $2R$, with its midpoint located at the origin and one mirror assembled 
to one of its ends, thus the latter is at fixed rigid distance $R$ from the origin. {\sl ii)}  A second mirror is free falling with its original position displaced slightly from the mirror attached to the bar, $x=R+\ell$. 

Because the mirrors are located at a rigid and geodesics points one should wonder which is the best set of coordinates to describe their dynamic. If one chooses the TT gauge we have the drawback that the mirror attached to the rigid bar will move after the interaction with the GW while the free falling mirror will remain at rest. If one instead chooses the LL the free falling mirror will oscillate after the interaction with the GW and the the rigid bar will be remain static. Any choice is plausible and leads to the same round trip delay time. We shall show the calculation in the LL gauge.

Because the photon bounces back and forth between the mirrors and only travels twice the bar length $R$ we calculate the delay time inside the $\ell$-cavity. Latter we could add the time delay related to the distance $R$. This last step will be unnecessary because, as we shall see, this delay is negligible.
 
Solving (\ref{photontrar}) under conditions (\ref{PHgencond}), with $\lambda\to t$, $z\to x$, $x_A=R$, ${x_B}= L + \frac{1}{2}h_+(t)L$ and $x_{B0}-x_{A0}=L-R=\ell$, one gets
\begin{eqnarray}
\label{delayIV}
\delta^{\rm SR} (t,\ell) = \left( {1 - \frac{{{\omega ^2}{R^2}}}{{2{c^2}}}} \right)\delta^{\rm LG} (t,\ell)+ \frac{R}{2c} \left[ h_+(t+{\frac{2 \ell}{c}}) + h_+(t) \right] \,.
\end{eqnarray}
Notice that (\ref{delayIV}) contains a term which is functionally equal to (\ref{delayI}) but with different parameters. Formally the main difference with (\ref{delayI}) is that (\ref{delayIV}) has two independent parameters, the rigid bar length $R$ and the initial separation between mirrors $\ell$. In the double limit $R\to 0$ and $\ell \to L$ (\ref{delayIV}) reduces to (\ref{delayI}). As we do not hasten to remark, if we add the time delay due to the $[0,R]$ strip Eq. (\ref{delayIV}) is gauge independent, thus independent of the coordinates.
\end{enumerate}

In what follows we shall elaborate on the differences between these three observables.

\section{Bouncing photons}

As mentioned earlier, we consider a gravitational perturbation with only the plus polarization and coming from the $z$-direction. 
If the interferometer is located in the plane $z=0$ and for a fixed GW frequency $\omega$ the perturbation takes the form 
\begin{equation}
\label{h}
h_+(t)= A \sin(\omega t)\,.
\end{equation}
With the previous expression we can calculate
the time delay after $n$-bounces for each of the observables (\ref{delayI}--\ref{delayIV}). We shall do so considering the perturbation in the $x$-direction and afterwards 
 add the effect in the $y$-direction discussing some modifications to the Michelson-Morley configuration. 

After $n$ bounces the time delay due to the mirrors is
\begin{equation}\label{nbouncing}
\delta^D(t,n,L)=\sum_{i=1}^n \delta(t+2(i-1)L/c,L)\,,
\end{equation}
with $\delta(t,L)$ given by (\ref{delayI}), (\ref{delayIII}) or (\ref{delayIV}) and the superscript $D\in ({\rm LG}, {\rm PR}, {\rm SR})$ will refer respectively to each of these situations.
\begin{enumerate}


\item[(LG)] {\bf Delay time for two free-falling mirrors}

This is the standard case. Inserting (\ref{h}) in (\ref{nbouncing}) and using (\ref{delayI}) one obtains 
\begin{eqnarray}
\label{delayIN}
 \delta^{\rm LG}(t,n,L)  &=& \frac{A}{{2\omega }}{\rm{ }}\left\{ {\cos (\omega t) - \cos \left( {\omega \left( {t + 2n\frac{L}{c}} \right)} \right)} \right\}\,.
\end{eqnarray}
This expression can be rearranged and after using some trigonometric relation we can rewrite it in terms of the ${\rm sinc}(x)$ function \cite{Maggiore:2008}
\begin{equation}
\label{dbook}
\delta^{\rm LG}(t,n,L)  = \frac{L n }{c } h_+\left(t+\frac{L  }{c }n\right)\, {\rm sinc}\left( \omega n\frac{L }{c } \right)\,.
\end{equation}

Because of the behavior of the ${\rm sinc}$ function, (\ref{dbook}) attains its maximum when $\omega\frac{L n }{c }\to 0$, from where one concludes that it is desirable that $L/c$ be much more smaller than the period of the GW,  in such a way that $\omega\frac{L n }{c }\ll 1$. If, contrariwise,   $\omega\frac{L n }{c }\gg 1$, during the traveling time $h_+(t)$ changes sign as many times that  multiple cancellations occurs and $\delta^{\rm LG}(t,n,L) $ is suppressed.


\item[ (PR)] {\bf Time delay for an optical bar}

This setup does not bring any new phenomena, exception of shedding some light to our
 procedure. 
 If we insert (\ref{h}) in  (\ref{nbouncing})  and using (\ref{delayIII}) one obtains 
 \begin{equation}
 \label{pprr}\hspace{-15mm}
\delta^{\rm PR}(t,n,L)  = \left(1-\frac{\omega L}{c}\csc\left(\frac{\omega L}{c}\right)\right) \delta^{\rm LG}(t,n,L)\,.
\end{equation}
As in the previous case we optimize to the maximum of the ${\rm sinc}(x)$ function. In this case the first non-vanishing terms in (\ref{pprr}) boils down to 
\begin{equation}
 \label{pprrnn}
\delta^{\rm PR}(t,n,L)= -\frac{n}{6}\frac{L^3\omega^2}{c^3}h_+(t) + {\cal O}(c^{-4})\,.
\end{equation}
With this result at hand it is not surprising that the authors in \cite{Braginsky:1997} did not find any competitive effect from this experimental setup. In order to increase the sensitivity one would have to increase substantially  the bar length bringing  new effects at play. On the other hand, this result justifies to neglect the $[0,R]$ strip in our calculations for the semi-rigid type device, $\delta^{\rm SR} (t,n,\ell)$ in the next section.


\item[ (SR)] {\bf Time delay for semi-rigid detector}

In the following we discuss the new configuration. With the use of  (\ref{h}) in  (\ref{nbouncing}) and taking into account (\ref{delayIV}) we get
\begin{eqnarray}
\label{delayIVN}
\hspace{-25mm} \delta^{\rm SR}(t,n,\ell)=  \left( {1 - \frac{{{\omega ^2}{R^2}}}{{2{c^2}}}} \right)\delta^{\rm LG}(t,n,\ell)
+ \frac{R}{c}{\cot}\left( {\omega \frac{\ell }{c}} \right){\sin}\left( {\omega  n\frac{\ell }{c}} \right) h_+\left( t +  n\frac{\ell }{c}  \right) .
\end{eqnarray}
The above expression contains a term that is proportional to $\delta^{\rm LG}(t,n,\ell)$ and follows the same pattern as (\ref{dbook}). Inside this term there is a contribution, $\frac{1}{2} \left(\omega \frac{ R }{c}\right)^2$, which can be safely neglected as ballpark is ${\cal O}(10^{-10})$. Thus nothings new so far.  The interesting term is the second one. As in the LIGO case but now with $\ell$ instead of $L$, we must fulfill $n\omega\ell/c\to 0$ in order to avoid the change of sign of $h_+$. Although we shall use the expression (\ref{delayIVN}) in our analysis it is instructive, in order to understand its behavior without resorting in any numerical evaluation, to study the leading term in $c^{-1}$ 
\begin{eqnarray}
\label{delaySR4}
\delta^{\rm SR}(t,n,\ell)&=&  \left(\ell +R \right)\frac{n}{c} h_+\left(t\right)\,.
\end{eqnarray}
Rewriting the first term as $\frac{1}{\omega} \left(\frac{\omega\ell\, n}{c}\right)$ we see that because we have look for to maximize the $\rm sinc$ function, i.e. $n\omega\ell/c\to 0$, the term between brackets is already small, thus the signal is very suppressed unless the gravitational source emits at very low-frequency. This will only increase the difficulty in obtaining and cleaning the data. This is not the case for the second term containing the rigid bar length $R$, that can be enhanced provided we properly chose the relevant scales for $\ell$ and $R$. That is, for a given $\omega$, whenever we can use an $\ell$ satisfying $n\omega\ell/c\ll 1$, $\delta^{\rm SR}$ increase linearly with $n$.
\end{enumerate}

Finally, we can answer the question raised in section 1 about the relativistic nature of the measurement. First we must clarify that by a non-relativistic measurement using light we understand that the mirrors trajectories follow the Newtonian equation of motion and that the speed of light is constant and greater than the relative  velocity between the mirrors. 
To first order in $c^{-1}$ we can rewrite
\begin{eqnarray}
\label{srnr}
\delta^{\rm LG}(t,n,L)\approx  2n \frac{1}{c}\,\, \frac{1}{2} L h_+\left(t\right)\,, \quad
\delta^{\rm SR}(t,n,\ell)\approx 2n \frac{1}{c}\,\, \frac{1}{2}  \left(\ell + R \right)h_+\left(t\right)\,,
\end{eqnarray} 
that can be interpreted as non-relativistic quantities. Contrariwise,  $\delta^{\rm PR}$ (\ref{pprr}) is clearly a relativistic effect due to the influence of gravitation on light travels.

In what follows we shall discuss $\delta^{\rm SR}$. So far we have not set any restriction to the lengths neither of the rigid bar, $R$, nor the distance between the two mirrors, $\ell$. In order to enhance the signal this will be our next task.


\section{Some preliminary constraints: a single arm detector}

Hitherto we have proposed three different experimental setups, Fig. 1 and Fig.~\ref{figSRLD},  from which two of them have already been constructed experimentally, Fig. 1 and  Fig.~\ref{figSRLD} (a). Let's see the main difference between an experimental setup as the LIGO configuration, Fig. 1, and our proposal, Fig.~\ref{figSRLD} (b). We focus on a single  arm geometry. In the former setup the Michelson-Morley interferometer length and the Fabry-Perot cavity length are identify. Its length, $L$, is optimized if it is approximately half of the GW length \cite{Press:1972am}. Contrariwise Fig.~\ref{figSRLD} (b) contains as independent quantities both: the length of the Michelson-Morley interferometer arm, $R+\ell$, and the size of the Fabry-Perot cavity, $\ell$, see Fig. \ref{interferometer}.

To substantiate the differences between the two setups let's remark two limiting cases in (\ref{delayIVN}):
\begin{enumerate}
\item[{\sl a})]  In the first case we recover the LIGO setup. If $  \ell \gg R\to 0\,,$ there is no substantial difference between the outcomes of (\ref{dbook}) and (\ref{delayIVN}) provided the length of the Michelson-Morley interferometer is of the order of the Fabry-Perot cavity. Comparing Fig. 1 and Fig.~\ref{figSRLD} (b) one sets $R\to 0$ and $L\sim \ell$. 
\item[{\sl b})] In the second case, reversing the order in the limits of the previous case, $ R \gg \ell\,,$ the last contribution to (\ref{delayIVN}) becomes leading with respect to the first one,
see (\ref{delaySR4}) and the arguments around it. This experimental arrangement has no parallel in (\ref{dbook}).
\end{enumerate}

Having at our disposal the three quantities $\{R,\ell\,,n\}$ one should wonder which is the optimal choice for this triad that results in an experimental apparatus of reasonable dimensions. To answer this question one has to take into account the current experimental limitations in different fronts. We review some of the shortcomings for each of the entries in the triad.  
\begin{enumerate}
 \item[{\bf $R$:}] Because the signal amplitude scales linearly with the rigid coordinate $R$ we would like to make this as large as possible while still considering the setup length small. If we increase the size we expect that the vibrational noise will increase and at some point will spoil the signal.  Notice that we do not resort in any resonant effect by considering the bars perfectly rigid. 

\item[{\bf $\ell$:}] If we reduce the Fabry-Perot cavity length below the $\mu$m or mm scale the thermal and the Brownian noise (respectively) would be dominant and one could not reach LIGO sensitivity. Thus these noises set the lower limit
to the length $\ell$ to the cm scale, see section \ref{nnoise}.
We have checked that for $1\,{\rm cm}<\ell<10\,{\rm cm}$ cavities and
 using the same quality mirrors as those in LIGO, the number of bounces may be increased due to the drastic decrease in the diffraction effect\footnote{ By \textit{diffraction effects} we mean those due to irregularities on the mirror surfaces and which do not lead in a loss of overall power, but rather in a redistribution of the laser intensity. }.

\item[{\bf $n$:}] Increasing the number of bounces, which seems to be a nice way out to increase the amplitude in the signal has the drawback of increasing the temperature in the mirrors. 
\end{enumerate}

Although we can set some lower limit for the Fabry-Perot cavity length using the impact of the noise contribution, the previous arguments indicate that 
the optimal triad can only be reach after a full simulation.  
We have refrained of doing an exhaustive analysis and instead
we have tested a full range of combinations in the triad that lead to the same conclusions. Here we show a representative point that
 works efficiently. More elaborated and refined ranges on $\{R,\ell\,,n\}$ will be given elsewhere\footnote{Work in progress.}. For the time being we take 
 \begin{equation}
 \label{input}
 R\approx 20\,{\rm m}\,,\quad  \ell\in [0.01,0.1]\,{\rm m}\,.
 \end{equation}
Bearing this range in mind, we briefly compare some similarities and differences
with LIGO (VIRGO) \cite{Maggiore:2008}, for which $L=4\times 10^3{\rm m}\,(L=3\times 10^3{\rm m})$. 
Comparing with LIGO (VIRGO), with finesse  ${\cal F}^{\rm LG}\approx 2\times10^2\, ({\cal F}^{\rm VG}\approx 50) $ the quality factor for the mirrors reduces up to $ Q \approx  10^{8}$ which translates in ${\cal F}^{\rm SR}\approx 22\times10^3$. It seems reasonable to reach $\approx 15\times 10^3$ bounces for this finesse. 
This leads into a significant decrease in the storage time $\tau_s$
\begin{eqnarray}
\label{estorage}
\hspace{-2cm}
\tau_s^{\rm LG}\approx \frac{L}{c} \frac{{\cal F^{\rm LG}}}{\pi} \approx \frac{4\times 10^3}{c}\, \frac{200}{\pi}\,
\approx 0.00084\,{\rm s}\,, \nonumber \\
\hspace{-2cm}
\tau_s^{\rm VG}\approx \frac{L}{c} \frac{{\cal F^{\rm VG}}}{\pi} \approx \frac{3\times 10^3}{c}\, \frac{50}{\pi}\,
\approx 0.00016\,{\rm s}\,, \nonumber \\
\hspace{-2cm}
\tau_s^{\rm SR}\approx \frac{\ell}{c} \frac{{\cal F^{\rm SR}}}{\pi} \approx \frac{\ell}{c}\, \frac{22\times 10^3}{\pi}\, \approx\left\{ \begin{matrix}
2.3\times 10^{-7}\,{\rm s};  & \ell=0.01\,{\rm m}\,, \\
2.3\times 10^{-6}\,{\rm s};& \ell=0.1\,{\rm m}\,,
\end{matrix}
\right.
\end{eqnarray}
that is, while $\tau_s^{\rm LG}$ is $\approx 10\%$ of the GW period, which has an expected frequency $\approx 100\rm Hz$,  $\tau_s^{\rm SR}$ is much more smaller. 

\begin{figure}
\centering
\includegraphics[scale=0.55]{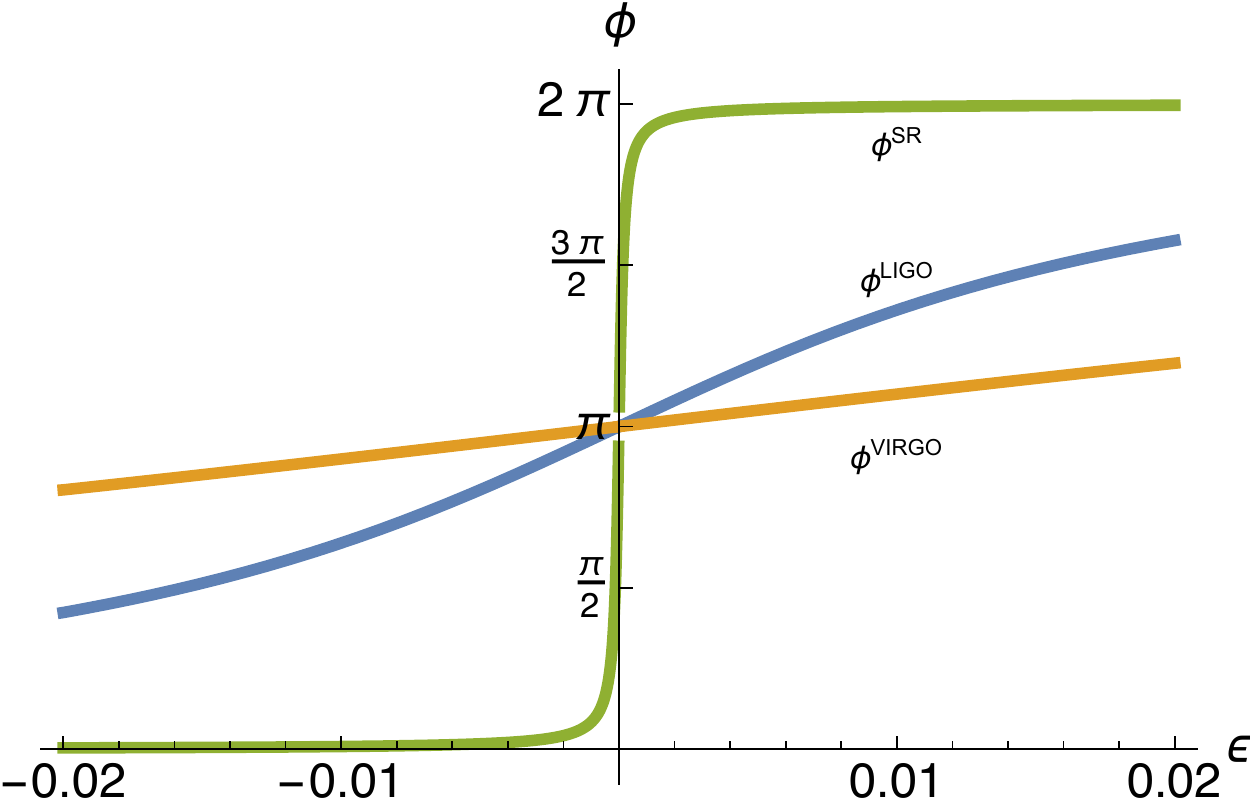}
\caption{Phase of the reflected field, as a function of $\epsilon$ for the
three different detectors: VIRGO, LIGO and SR.}
\label{phi}
\end{figure}

It is clear that because the SR setup allows a larger finesse this translates directly in an increase of the coupling rate $\sigma$
\begin{equation}
\sigma = \frac{p{\cal F}}{\pi}\,,
\end{equation}
being $p$ the losses in the first mirror. Typical values in VIRGO and LIGO are $p\sim 2\times 10^{-5}$. Using this we get
\begin{equation}
\sigma^{\rm LG}\approx 10^{-3}\,,\quad  \sigma^{\rm VG}\approx 10^{-4}\ \,,\quad  \sigma ^{\rm SR}\approx 0.1\,.
\end{equation}
LIGO and VIRGO cavities are well overcoupled, $0< \sigma < 1$, but  the SR cavity is more near optimality, $\sigma=1$. This leads to a higher sensitivity to changes in the phase, $\phi$, of the reflected field from the mirrors  
\begin{equation}\label{philength}
\phi =\pi+{\rm arctan} \left(\frac{{\cal F} \epsilon}{\pi} \frac{1}{1-\sigma}\right) + 
{\rm arctan} \left(\frac{{\cal F}\epsilon}{\pi} \right)\,,\quad \epsilon=\frac{4\pi \Delta {\rm length} }{\lambda_0}\,({\rm mod}\, \pm 2\pi)
\footnote{{\sl length} refers to the cavity length, $L$ for LIGO (VIRGO) or $\ell$ for SR.} \,,
\end{equation}
where $\lambda_0$ is the laser wave length. In the absence of GW, $\epsilon\approx 0$, the cavity is near the resonance peaks of the Fabry-Perot.  In Fig. \ref{phi} we have plotted the phase of the reflected field, as a function of $\epsilon$, for the three different detectors. 

\section{Signal Analysis}
\label{Sign_Anal}

To pursue further our analysis we need a few experimental inputs from the LIGO detector. The interferometer main characteristics are its arm length $ L\approx 4\,$km and the laser wave length $\lambda_0= 1064\,$nm  \cite{Abbott:2016blz}. Each laser beam bounces back and forth about $n=280$ times before they are merged together again \cite{ligopage}. The first black hole detected had an orbital frequency of $f = 75\,$Hz, half of the GW frequency.
Bearing these inputs in mind let's analyze the consequences of (\ref{dbook}) and (\ref{delayIVN}). 

First we have checked numerically the correctness of (\ref{delaySR4}): 
choosing $\ell$ small enough, the first term in (\ref{delayIVN}) is negligible while the second is enhanced. In this case the second term is almost insensitive to the frequency of the signal and within a wide range of values it increases directly with number of bounces. 
As far as we are concerned, this result is new and represents a qualitative leap in the study of GW detectors.

\begin{figure}
\centering
\subfigure[]{
\includegraphics[scale=0.23]{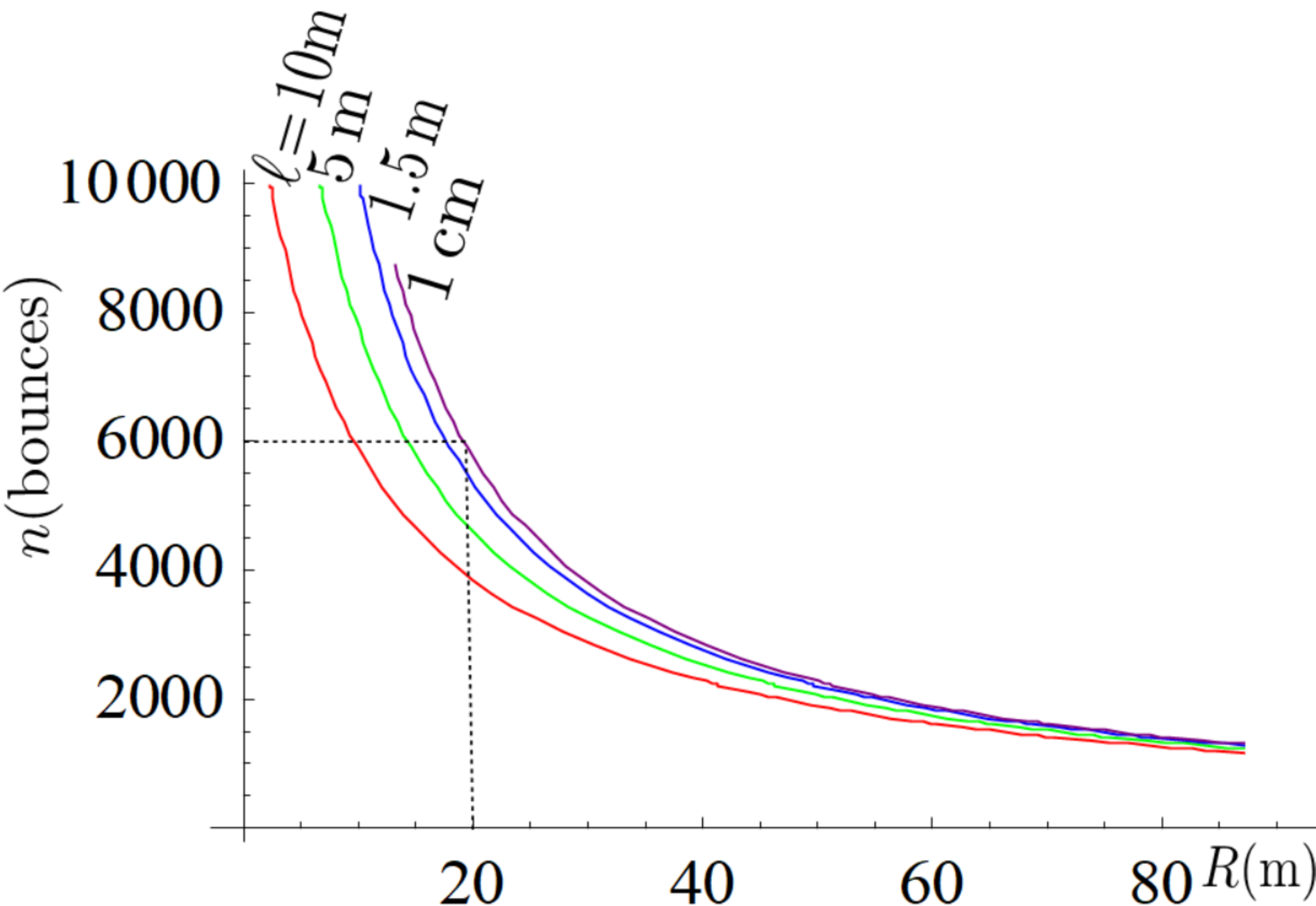}}
\subfigure[]{
\includegraphics[scale=0.27]{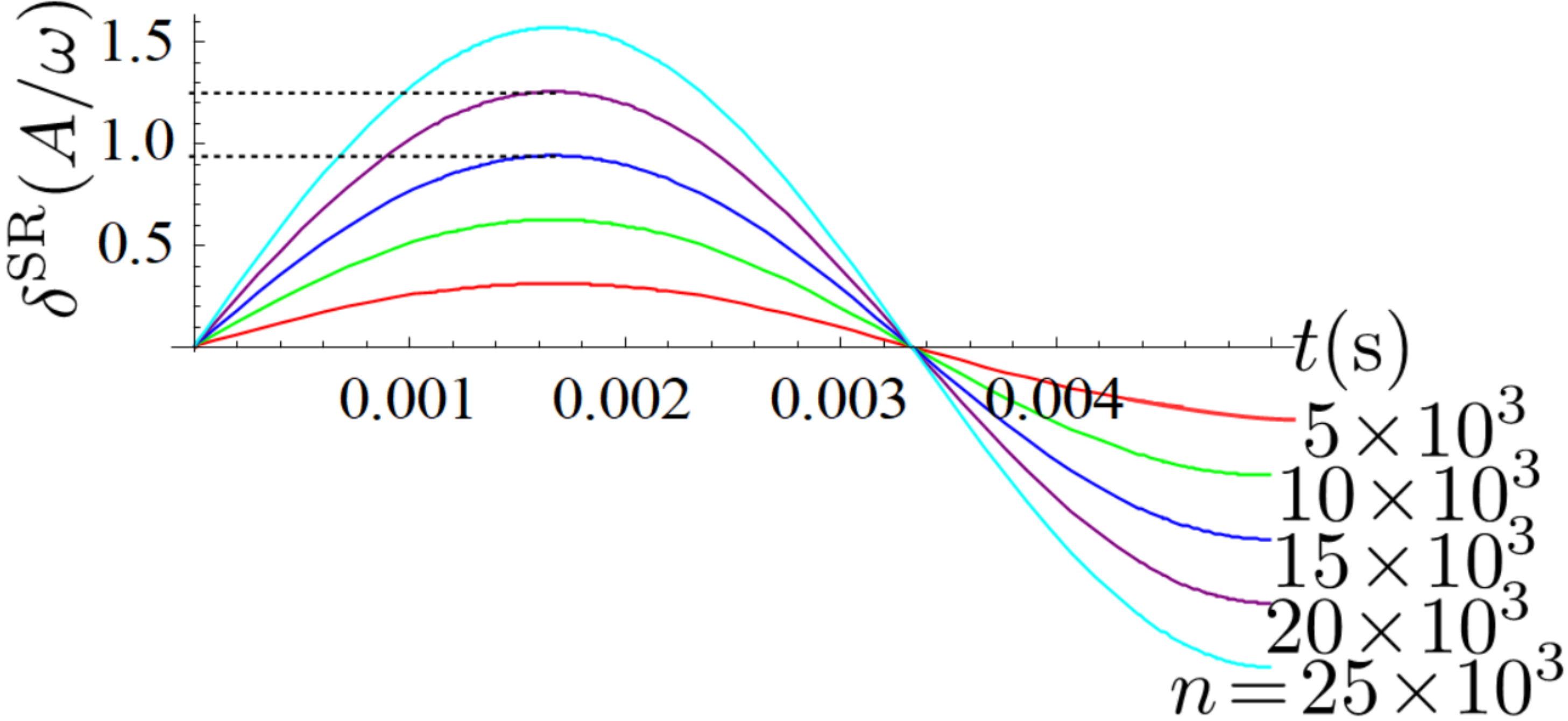}}
\caption{(a)  Number of bounces  needed in (\ref{delayIVN}) to match the results of (\ref{delayIN}), using LIGO data ($\delta^{\rm LG}_{\rm max} \approx 0.36\, A/\omega\,$),  as a function of the mirror separation $\ell$ and the rigid coordinate $R$. (b) The time delay (\ref{delayIVN}) increases significantly with the number of bounces. Here we have set $R=20\,\rm m$ and $\ell=1\,\rm cm$.}
\label{Nboun}
\end{figure}

We have looked for the number of bounces needed in (\ref{delayIVN}) as a function of the bar length, $R$, and the separation between the mirrors plates, $\ell$, to match the outcome of (\ref{dbook}) with LIGO data,  $\delta^{\rm LG}_{\rm max} \approx 0.36\, A/\omega\,$. We show in Fig.~\ref{Nboun}-(a) this dependence for different choices of the parameters $\{\ell,R\}$. As is evident, for a fixed $\ell$, a reduction on the size of the rigid component has to be compensated with an increasing number of bounces to keep the amplitude in the signal. 
For a relatively short arms lengths $R \approx 20\,$m one can reduce the Fabry-Perot cavity to less than $1\,$m and still get the very same results as LIGO but with a few thousand bounces instead of two hundred. 
This could be done because the Fabry-Perot cavity length can be reduced at will. Thus for a fixed number of bounces one can fine tuning the values of $\{\ell,R\}$ and obtain, to the least, the same amplitude in the signal as LIGO. Can we improve this? As is shown in Fig.~\ref{Nboun}-(b) for a fixed pair $\{\ell,R\}$ one can increase the number of bounces and surpass the LIGO signal amplitude for a factor $5$ or more. The reason is, as mentioned above, that Eq. (\ref{dbook}) has a very mild dependence in the number of bounces, while (\ref{delayIVN}) is linear in these, see (\ref{delaySR4}) and arguments around it. As mentioned above, section 6, the number of bounces should have an upper limit due to the thermal increase in the mirrors.

One should wonder what happens if we simply approach LIGO's mirror and increase the number of bounces in the process. First and foremost in this case $\delta^{\rm LG}$ has no reference to the rigid coordinate and thus approaching the mirror decrease the amplitude of the signal. For separations around  $L \approx 20\,$m we need over $6000$ bounces to obtain $\delta^{\rm LG}_{\rm max}\,$. But for a $20\,$m-cavity and for such a number of bounces, diffraction effects are still dominant over reflection. This translates on a quality factor $Q \approx  10^{11}$ and on a maximum number of bounces around a half the minimum needed.
This stress the difference between  (\ref{dbook}) and (\ref{delayIVN}). The former needs large distances between the mirrors because bouncing is not  enough additive while the latter is more additive and, more important, we can skip the diffraction effect.

Summing up our findings, (\ref{delayIVN}) contains a contribution that is enhanced with an increasing number of bounces.  This depends mainly on the Fabry-Perot cavity length $\ell$ that is not related with the interferometer arm length $R+\ell$ and can be made as small as desired. We explore next this possibility.


\section{A possible experimental setup}

\begin{figure}
\centering
\subfigure[]{
\includegraphics[scale=0.3]{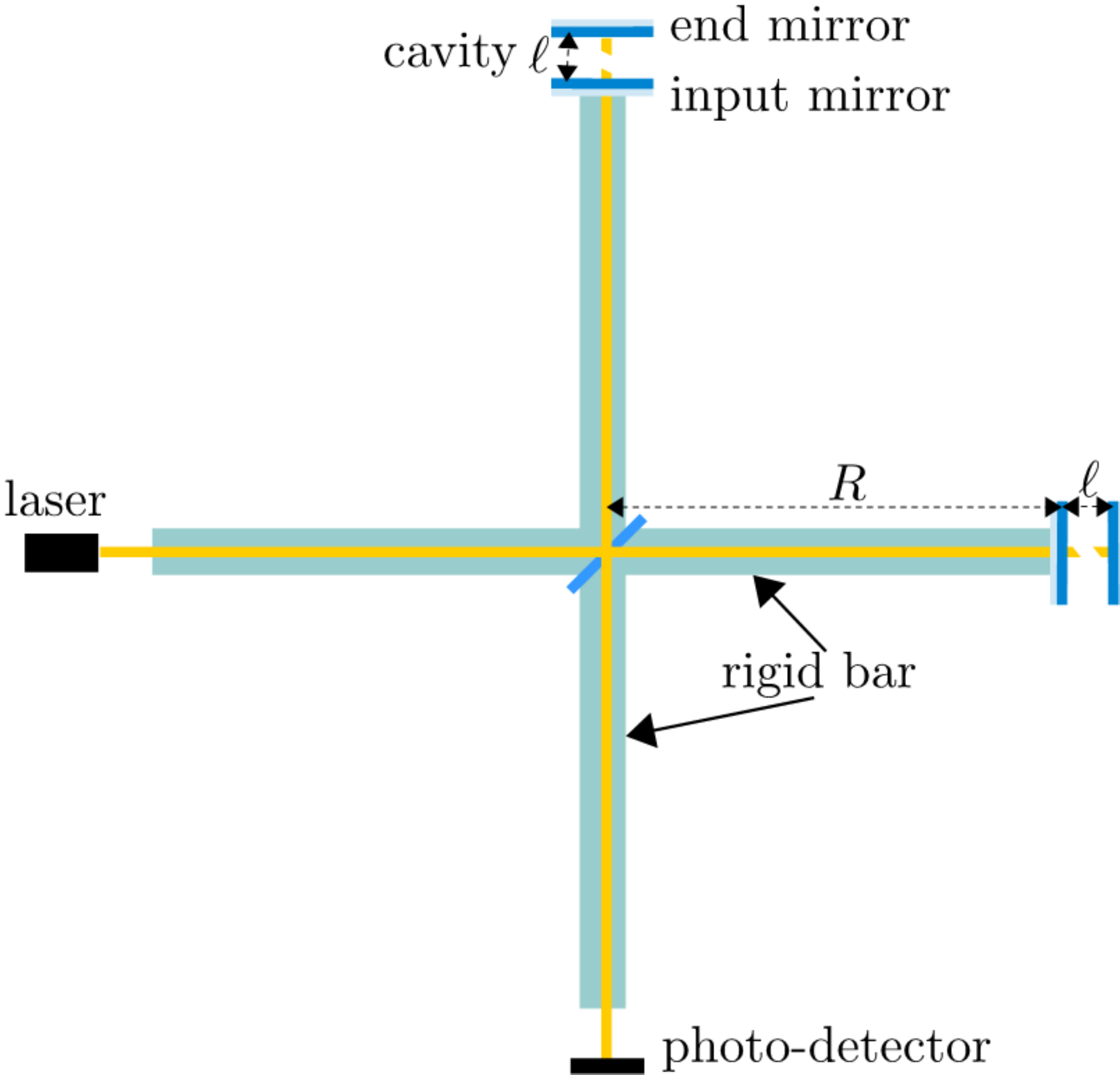}}
\subfigure[]{
\includegraphics[scale=0.2]{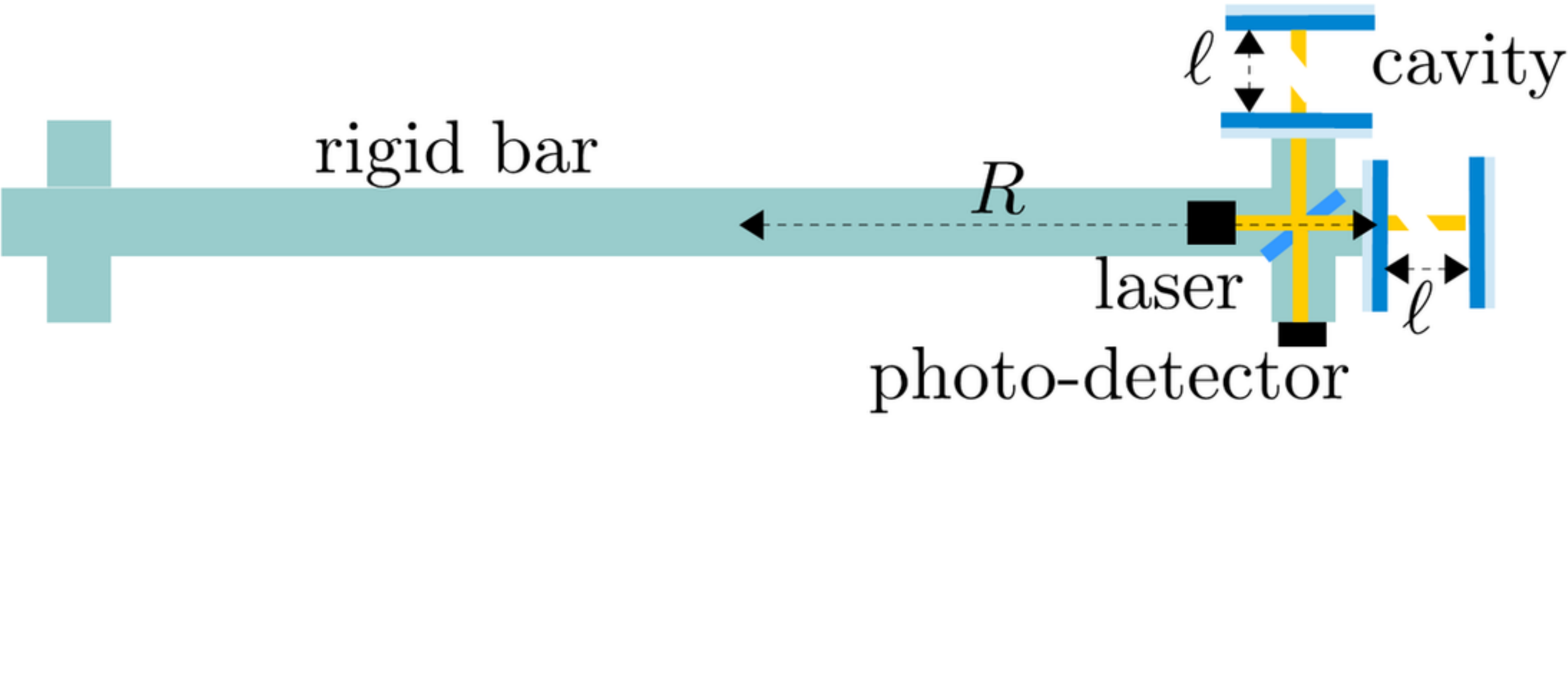}}
\caption{Detectors optical configurations. The bars are quasi-rigid objects. Elements are not scaled.}
\label{interferometer}
\end{figure}

In Fig. \ref{interferometer} we present two possible experimental designs.  The general disposition of both 
mimics essentially that of LIGO. They consists in a Michelson-Morley interferometer with two Fabry-Perot cavities. The two input mirrors are attached to the rigid bars' edges while the pair of end mass-mirrors are in free-fall. In the first case, Fig. \ref{interferometer} (a), the full apparatus consists of a mixture of a two orthogonal ``Weber like'' resonant bars detectors \cite{weber} couple to the Fabry-Perot interferometer that amplifies the signal. Notice that, concerning the bars, we do not resort in any resonant effect by considering the bars perfectly rigid. Nevertheless, it is expected that, by adapting the  bar design to the GW frequency, the motion of the input mirrors may be in opposite phase with respect to the end mirrors. The second possibility, Fig. \ref{interferometer} (b), uses an asymmetric  Michelson-Morley interferometer. At first sight, with configuration (a) we get twice the performance of (b). In both cases we could incorporate the photo-detector and the laser on top of the semi-rigid bars, thus reducing the misalignment of all degrees of freedom \cite{aligment}.

The dimension of either design is that of the bar, $2R$, which, as discussed above, is relatively small. So we can think of a space experiment. This will have a threefold consequence: 
\begin{enumerate}
\item[{\sl i)}] Minimize the noise due to the motion of the mirrors from ground vibrations. 
\item[{\sl ii)}] Minimize the interactions of the residual gas particles in between the mirrors and the laser light, pressure is $\sim 10^{-11}$Pa. The latter is also reduced because the distance between mirrors, $\ell$, can be made relatively small. 
\item[{\sl iii)}] Although temperature drops, around $50\,\rm K$, with respect to the room temperature there is experimental evidence that this does not lead to a substantial improvement of the thermal noise \cite{Franc:2009nb}. 
\end{enumerate}
Although a space based experiment possibility exists, a system of suspended mirrors and arms on the earth can not be ruled out on the sole base of vibrational noise \cite{vibration}.

We shall focus on the sequel in the design shown in  Fig. \ref{interferometer}-(a) and find the phase difference in the Fabry-Perot component
of the detector.

{\bf Fabry-Perot:} due to its location, the motion of the free mirror of the Fabry-Perot interferometer depends on $R$. That is, in (\ref{philength}) for LIGO (VIRGO) $\Delta{\rm length}=\Delta L=\frac{L}{2} h_+$ but for SR $\Delta{\rm length}=\Delta \ell=\frac{R+\ell}{2} h_+$. The result for the phase difference due to the Fabry-Perot interferometer after $n$-bounces can be read out from the standard analysis, \cite{Maggiore:2008}:
\begin{equation}
\label{phaseFP}
\vert \Delta \phi_{FP}\vert = A \frac{4{\cal F}}{\pi} k_0\, {\rm length}\, \frac{1}{\sqrt{1+\left(f_{gw}/f_p\right)^2}}\,,
\end{equation}
where now {\sl length} refers to the location of the second mirror, $L$ for LIGO (VIRGO) or $R+\ell$ for SR, $f_{gw}=\frac{\omega}{2\pi}$ and the pole frequency, $f_p$, is just the normalized inverse storage time (\ref{estorage})
\begin{equation}
\label{fs}
f_p=\frac{1}{4\pi\tau_s}
\approx\left\{ \begin{matrix}
93\,{\rm s}^{-1};  & {\rm LIGO}\,, \\
500\,{\rm s}^{-1};  & {\rm VIRGO}\,, \\
34090 \,{\rm s}^{-1};& {\rm SR}\,\, \ell=0.1\,{\rm m}\,,
\end{matrix}
\right.
\end{equation}
Notice that in the case of the SR detector, $f_p\gg f_{gw}$, and the factor with the squared root is just equal to the identity.

\begin{figure}
\centering
\subfigure[]{
\includegraphics[scale=0.6]{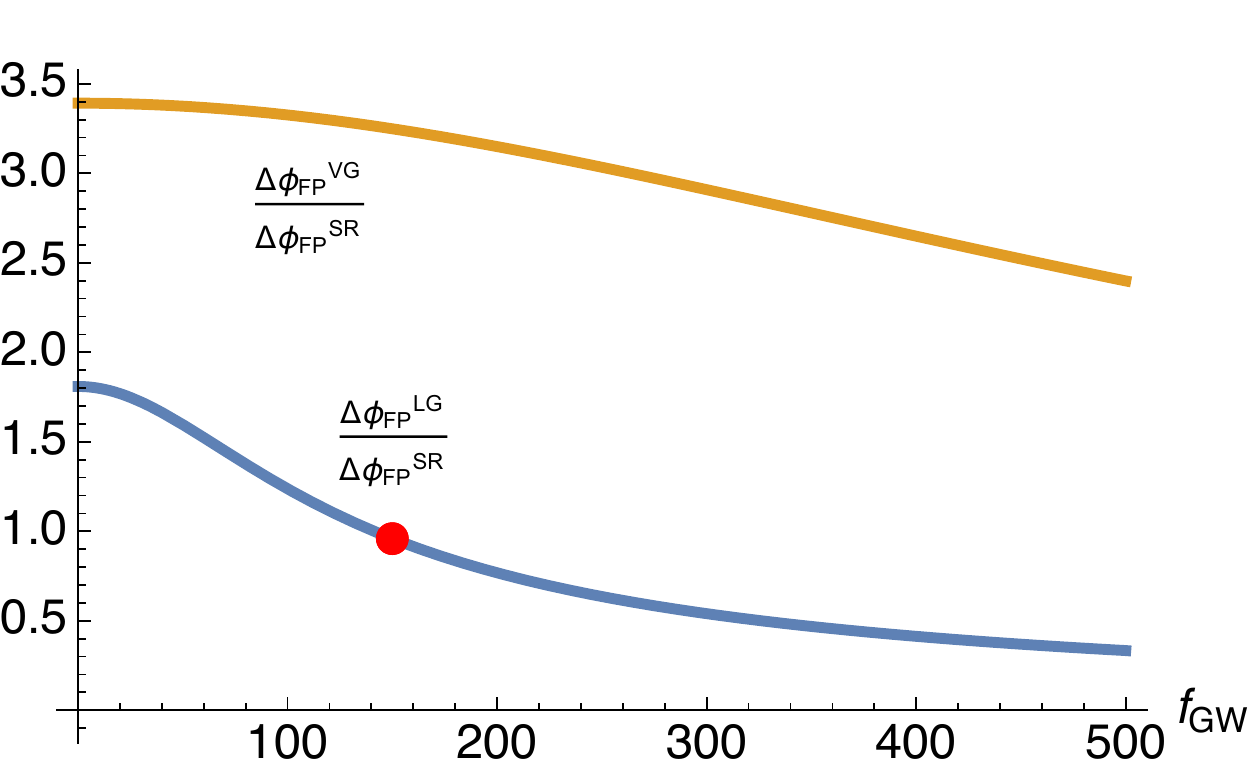}}
\caption{The ratios of phase shift in the VIRGO \& LIGO Fabry-Perot interferometer with respect to SR Fabry-Perot interferometer as a function of the GW frequency. The SR detector has $\ell=10\,{\rm cm}$. For comparison purposes we have shown with a dot the frequency of the first detected GW event.}
\label{pow}
\end{figure}

In Fig. \ref{pow} we show the ratio $\frac{\vert \Delta \phi_{FP}^{LG}\vert}{\vert \Delta \phi_{FP}^{\rm SR}\vert}$ 
and $\frac{\vert \Delta \phi_{FP}^{VG}\vert}{\vert \Delta \phi_{FP}^{\rm SR}\vert}$  as 
function of the GW frequency $f_{gw}$.  For the SR detector we used a cavity with $\ell=10$ cm. As is already evident in (\ref{phaseFP}) if the phase difference in the new experimental setup must not deviated significantly from the LIGO result the reduction in the length of the interferometer cavity must be  compensate by the increase in the finesse. This is indeed the case.

Although our results are very crude and still lack many factors, confronting (\ref{phaseFP}) with the performance of current detectors reveals that both the phase difference inside the Fabry-Perot cavity can give competitive results. To put it different, is the fact that we measure with respect to the end of the rigid bar what allows to
obtain an increase in the signal while reducing  the size of the detector.


\section{A noisy zoo}
\label{nnoise}

Having defined the experimental setup and although the previous analysis looks promising for cavities $1\,{\rm cm}\lesssim \ell\lesssim 10\,$cm, gravitational signals are rather weak and turns to be masked by a plethora of background. Our aim is to analyze approximately their main contributions and how it will affect and restrict the interferometer design. 

As a preliminary task we refer to the mirrors.

\noindent {\bf Mirror characteristic:}  the mass of the mirrors is $M=20\,{\rm kg}$. As simplification we take both faces, HR and AR coating, to be identical and assume a single coating layer with a round trip loss equal to that at room temperature, less than $50\,\rm ppm$ \cite{Pinard}. 
The new ingredient that we have to take into account in the analysis of the new configuration is that there is an increase in temperature of the mirror surface due to the increasing number of photon bounces.
We take the worst case scenario, the so called in-plane heat transport, where all the absorbed energy goes to increase the temperature in the coating without transferring heat to the substrate.  This can be justified, in the LIGO case, since the apparent thermal conductivity for the coating is higher than for the substrate. Inside the range of temperatures in which the apparatus operates, we find that the elastic modulus and thermal properties of the, un-doped, $Ta_2O_5$ has only tiny variations \cite{ta2o5,ta2o5bis}. 

{\bf A noisy zoo:} although we present this analysis at the end, the path we have followed is the inverse: once we realized that we have a contribution that can increase the signal in the detector, to set up the optimal design we
analyzed the noise contribution. This analysis naturally bounds the length and finesse of the Fabry-Perot cavity but leads the length of the Michelson-Morley length free. As we shall see there is a nice interplay between them in the noise analysis when fixing the total length of the interferometer.

To start with, let's fix the distance between the mirrors in the Fabry-Perot cavity and rule out, on the sole basis of noise, some scales. The ballpark of the noise in the LIGO case is driven, in order of importance, by the Brownian, coating thermoelastic and  thermo-refractive noise. In order to compare the performance of the SR detector we take the former of the mentioned noises as our bench mark reference and work out the conditions under which the Fabry-Perot interferometer in the SR detector reach an error
of the same order as this.
\begin{enumerate}
\item[{i)}] If we use a micro-cavity the mirrors must be treated as curved. As far we are aware of only the thermo-refractive noise have been calculated for such geometry and for fussed silica. In this setup the thermal fluctuation around frequencies $f\sim 10^2\,$Hz is a factor $10^{11}$ bigger than the required \cite{thermalsphere}. 
\item[{ii)}] If we increase the size of the cavity to the $\rm mm$ scale the thermo-refractive noise decreases at the expenses of increasing the Brownian noise, which turns to be once more the main contribution. Even thought still we are several orders of magnitude far from our bench mark point \cite{cole}. 
\item[{iii)}] The previous two scale cavities are ruled out because of the noise. We find then mandatory to increase the cavity length up to the $\rm cm$ scale. At this scale the mirrors distance is $10^4$ bigger than the laser wave-length and we can use with confidence the same expressions for the noise as in LIGO, Table \ref{noise}.  With arms length of $L=20\,$m ($L=40\,$m) and distances $\ell\sim 1$ cm between the Fabry-Perot mirrors we need over $6000$ ($3000$) bounces to match LIGO results. This leads to, for both cases, a quality factor $Q\approx 10^8$, which translates in a finesse ${\cal F}\approx 22 \times 10^3$, which is attainable with commercial cavities \cite{cav}. 
\item[{iv)}] One should wonder why not to increase even more the scale. If we do so the diffraction effects wash out the signal.
\end{enumerate}

Thus if the cavity length is $1\,{\rm cm}\,\lesssim \ell \lesssim 10\,$cm there is some room to obtain a sensitivity similar to that in LIGO. To ascertain, at least partially, the viability of this statement we have to quantify the departure from the actual noise in LIGO. For that we defined the {\it{relative deviations}}
\begin{equation}
E= \frac{1}{N} \sum_i X_i\,,\quad X_i:= \frac{S_i}{S_i ^{\rm LG}}\,,
\end{equation}
where $i$ runs over all possible sources of noise, see Table \ref{noise}\footnote{As we are not original in this piece of work we refrain to explicitly explain all the ingredients in the table and refer the reader to the reference.}. For a number of bounces of $6000$ the deviation due to the increase in temperature at the mirror surface is less than $E < 1\%$ which still support our statement. 

To conclude this analysis we pay some attention to the laser shot noise and the radiation pressure on the mirrors.
In a wide range of optical systems these two noises are considered as the main limitation factors of sensitivity \cite{thermal}. For reference purposes we use the same wave-length laser as LIGO. The combination of the two noises gives us the strain density for the optical read-out noise
\begin{equation}
S_n(f)_{\rm opt}= S_n(f)_{\rm shot}+S_n(f)_{\rm rad}\,.
\end{equation}
The shot noise can be read directly from the one in the Michelson-Morley interferometer
\begin{eqnarray}
S_n^{1/2}(f_{gw})_{\rm shot}&=& \frac{1}{8 {\cal F}\, {\rm length}} \left(\frac{4\pi\hbar\lambda_0 c}{\eta P_{bs}} 
\right)^{1/2} {\sqrt{1+\left(f_{gw}/f_p\right)^2}}\,,
\end{eqnarray}
where {\sl length} stands either for $L$ in the case of LIGO (VIRGO) or $\ell+R$ in the case of the SR detector and, as is customary, $\eta$ is the reduction factor in the effective power of the photodiode, $\eta\sim 0.93$. $P_{bs}$ is the power on the beam-splitter after recycling, $P_{bs}=C P_0$ with $C\sim {\cal O}(10)-{\cal O}(10^2)$ and $P_0=20W$.

The radiation pressure part needs a slightly more rearrangement. Taking the spectral density of the displacement of the  mirror we must divide it by the transfer function that relates $\Delta {\rm lenght}$ to the GW amplitude $A$. This can be read out, in our case, directly from (\ref{srnr}). Thus the transfer function is simply $\ell+R$. Introducing the amplification due to the Fabry-Perot and the factors due to the second Michelson-Morley interferometer arm we arrive at
\begin{eqnarray}
S_n^{1/2}(f_{gw})_{\rm rad}&=& \frac{16\sqrt{2} {\cal F}}{M\, {\rm length}\, (2\pi f_{gw})^2}\left(\frac{\hbar}{2\pi} 
\frac{P_{bs}}{\lambda_0 c}\right)^{1/2} \frac{1}{\sqrt{1+\left(f_{gw}/f_p\right)^2}}\,, 
\end{eqnarray}
where {\sl length} has the same meaning as before.

 \begin{figure}
\centering
\includegraphics[scale=0.7]{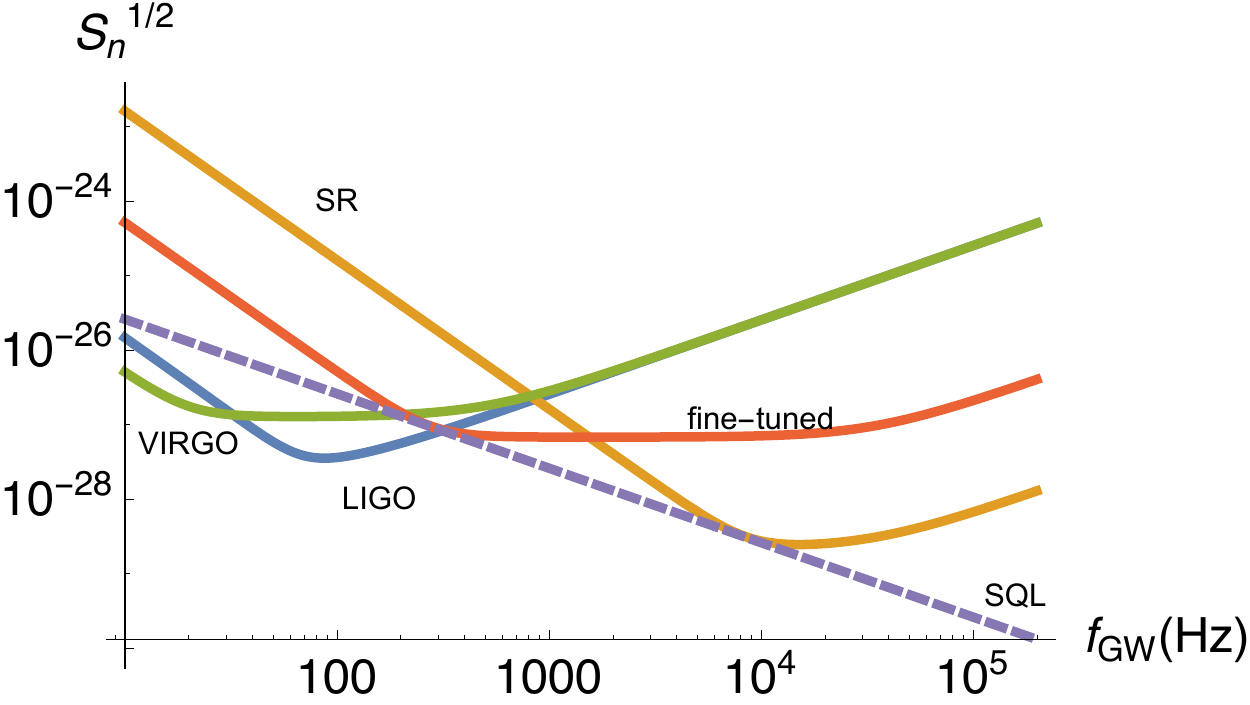}
\caption{ The strain sensitivity of the optical read-out-noise compared with the SQL pseudospectral density. We have used numerical values of the various parameters described in the text. The fine tuning curve stands for a recycling factor $f_0\approx 280$.} 
\label{noisee}
\end{figure}

In Fig. \ref{noisee} we show the strain sensitivity of the optical read-out noise. It's clear that, in this very crude analysis, with the previous inputs the SR performance is better at high frequencies, $f_{gw}\sim 800$ Hz, but is not competitive at relative low-frequencies $\sim 100$ Hz. 
The main contribution for frequencies $\ll 800$ Hz is the radiation pressure. In this region of frequencies the interplay between the several factors is given by 
\begin{equation}
f_0= \frac{8 {\cal F}}{2\pi} \left(\frac{P_{bs}}{\pi\lambda_0 c \, M}\right)^{1/2}\,.
\end{equation} 
Thus in order to decrease the right hand side of the plot one of options is to decrease the power in the beam-splitter. A little of fine tuning shows that for $f_0\approx 280$ Hz we can reach a competitive behavior above Medium Frequency events $f_{gw}\approx  300$ Hz, this is just a factor two bigger than the first recorder GW. 
Between the phenomena available to the range of frequencies of the SR detector  we can find emitters as black- holes up to $10^2 M_{\odot}$, collapse of stars, Weber bursts and Supernovae  \cite{Press:1972am}. For comparison purposes we also shown the Standard Quantum Limit (SQL) strain sensitivity
\begin{equation}
S^{1/2}_{SQL}(f_{gw})=\frac{1}{2\pi {\rm length}} \sqrt{\frac{8\hbar}{M}}\,.
\end{equation}


\begingroup
\begin{table*}
\makegapedcells
\setlength\tabcolsep{6pt}
    \begin{tabularx}{\linewidth}{c X}
   \Xhline{0.8pt}
Noise(s)     &   $S_i$     \\
   \hline
Substrate Brownian & 
$\frac{2 k_B T \phi(f)(1-\sigma_s^2)}{\pi^{3/2} Y_s \omega f}$
 \\
 Coating Brownian & 
$\frac{2 k_B T \phi_{\rm coat}(f)(1-\sigma_s^2)}{\pi^{3/2} Y_s \omega f}$
 \\
Substrate thermo-elastic&
$\frac{4 k_B T^2 \alpha_s (1-\sigma_s^2)^2 \kappa_s}{\pi^{5/2} (C_s \rho_s)^2 \omega^3 f^2}$ 
\\
Coating thermo-elastic&
$\frac{8 k_B T^2(1-\sigma_s^2) \alpha_c^2 d_N^2}{\pi^{3/2} \sqrt{\kappa_s C_s \rho_s}\omega^2 f^{1/2}}
G^{\rm coat}_{\rm TE}(\omega)$ 
\\
Coating thermo-refractive&
$\frac{2 k_B T^2 \beta_{\rm eff}^2 \lambda^2}{\pi^{3/2} \sqrt{\kappa_s C_s \rho_s}\omega^2 f^{1/2}}$ 
\\
Substrate photo-thermo-elastic&
$\frac{\alpha^2 S_{\rm abs}}{\pi^4 C_s^2 \rho_s^2 \omega^4 f^2}$ 
\\
Coating photo-thermo-elastic &
$\frac{4 (1+\sigma_s)^2 S_{\rm abs}\alpha_c^2 d_N^2}{\pi^3 C_s \rho_s \kappa_s \omega^4 f}
G_{\rm surf}^{\rm coat}(\omega)
$ 
 \\
 Coating photo-thermo-refractive&
$\frac{ S_{\rm abs}\beta_{\rm eff}^2 \lambda^2}{\pi^3 C_s \rho_s \kappa_s \omega^4 f}
G_{\rm surf}^{\rm coat}(\omega)
$ 
 \\
Substrate and coating Stefan-Boltzmann thermo-radiation&
$8 \sigma_B k_B T^5 \pi \omega^2$ 
\\
\Xhline{0.8pt}
\end{tabularx}
\caption{\label{noise}Possible noise contribution to the coating and substrate of optical mirrors, see \cite{thermal} for details.}
\end{table*}
\endgroup


\section{Remarks}
We have started by a simple exercise suggesting that, from the detection point of view, the phenomenon of GW could be cast into the Newtonian theory. We have studied two gauges especially suitable for the case we are interested in: the Gaussian and the Fermi-rigid. In the definition of the Fermi Rigid gauge we have used the Rigid gauge that intent to define a set of space coordinates $x^i$ that can identified as the points attached to a suitable rigid body. 

We have applied these findings to the plus mode of a linear plane GW. In this context the GG gauge is restricted to the transverse traceless gauge while the FR gauge coincides with the local Lorentz gauge.  Although the result for the LL gauge near the detection zone is well known, we gave a rationale to obtain it using the RG gauge. 

Next we have found the  time delay for a photon bouncing once between mirrors in three different setups. Two of them are already known, the LIGO like device and the optical bar device, and we find coherent results. The third setup, the semi-rigid like device,  is as far as we know, new. We looked at the $n$-bounces time delay for the three setups. These delays constitute the observables that the devices are trying to measure. With respect to the question about the relativistic character of this observables, we have confirmed that the leading terms of both, the LIGO and the semi-rigid like devices, have a non-relativistic character.

What is really novel is that, although the signal in LIGO and the semi-rigid devices grows linearly  with the number of bounces, in the latter the length of the cavity $\ell$ does not restrict  the amplitude of the motion of the mirror, which depends on the length of the rigid bar $R$. That is, we can drastically decrease the length of the cavity thus increasing the number of bounces without decreasing the amplitude of the relative motions of the mirrors.

We have analyzed  the  feasibility of the semi-rigid like device  from the experimental point of view. By choosing appropriate scales for the length of the bar and the cavity the semi-rigid setup can be competitive with LIGO for GW frequencies larger than $300$ Hz. On the other hand, it even seems that the dimensions of the device allow it to operate in outer space.

Although we have not modeled the full gravitational detector, we have checked that the main parts, quasi-rigid bars and optics, do not lead to unexpected and large noises. 

In conclusion we have paved an interesting way to relook gravitational wave detectors. The approach only rely on an extensive use of the concept of rigidity. Despite the controversy that generates the adaptation of this concept to GR, these results seem clear enough to seriously consider their use at a practical level.

\section*{Acknowledgments}
We would like to thank to R. Herrero for discussions on optical mirror systems, to J. Llosa  and R.Torres  for a careful and critic reading of the manuscript. And finally to the anonimous referees which initial questions have led us to improve the article to the current version. P.T. is partially supported by MINECO grant FPA2016-76005-C2-1-P.

\end{document}